\documentclass{amsart}
\usepackage{amsmath}
\usepackage{mathtools}
\usepackage{amssymb}
\usepackage{mathrsfs}
\usepackage{bbm}
\usepackage{dsfont}

\begin{document}

\title{Emergence of Kastor-Traschen Spacetime from a Nonlinear Dirac System}

\author{Daisuke Ida}
\email{daisuke.ida@gakushuin.ac.jp}
\address{Department of Physics, Gakushuin University, Tokyo 171-8588, Japan}

\date{25th Aug. 2026}

\begin{abstract}
  We establish a uniqueness theorem in four-dimensional spacetime,
  demonstrating that a nonlinear spinor system naturally constrains the background geometry to the Kastor-Traschen spacetime,
  which describes dynamical, cosmic multi-black-holes.
  By analyzing a nonlinear field equation where the super-covariant derivative of a Dirac spinor is sourced by intrinsic fermion currents, we evaluate its integrability under two physical requirements: a phase-locking condition enforcing a real hermitian product \(\overline{\psi}_R \psi_L\), and a pure-electric source condition.
  We show that the field equation uniquely compels the spinor to be an eigenvector of the super-covariant Dirac operator, with the eigenvalue dynamically identified as the Hubble constant of the de Sitter background.
  These results provide a novel geometric mechanism wherein both the cosmological constant and dynamical black hole configurations spontaneously emerge from fermion condensates.
\end{abstract}

\maketitle

\section{Introduction}
The interplay between fermionic degrees of freedom and spacetime geometry plays a central role in supergravity and quantum gravity. A classic result in this direction is Tod's uniqueness theorem~\cite{GH82,Tod83}, which demonstrated that the existence of a super-covariantly constant spinor in four dimensions freezes the background geometry into the stationary Israel-Wilson-Perj\'es spacetimes~\cite{Per71,IW72}. In the static case, these reduce to the Majumdar-Papapetrou spacetimes describing extreme balanced black holes~\cite{HH72}. Extending this framework to a dynamical background with a positive cosmological constant ($\varLambda > 0$), namely the Kastor-Traschen spacetime, presents a more challenging problem. The Kastor-Traschen geometry describes coalescing, extremal black holes embedded in the de Sitter universe~\cite{KK93}, yet its microstructural origin from a fundamental spinor field equation has remained unclear.

In this article, we close this gap by proving a uniqueness theorem for the Kastor-Traschen geometry. Our key insight is that tracking dynamical cosmic evolution requires accounting for nonlinear feedback from the fermion field. Rather than imposing a conventional linear equation, we investigate a nonlinear Dirac system where the standard super-covariant derivative $D_\mu$ is naturally deformed by a chiral-covariant
fermion self-interaction:
\begin{align}\label{eq:scc}
  D_\mu\psi = \frac{H}{2}\gamma_\mu\psi - \frac{H}{4W^2}
  (\bar{\psi} \gamma_\mu\psi)(\bar{\psi}\gamma_\nu\psi)\gamma^\nu\psi,
\end{align}
where $H$ is a real constant parameter representing the scale of the cosmic expansion, and the scalar $W:=\overline\psi_R\psi_L$ is the hermitian product of the chiral components. 

The main discovery of this work is that when this system is subjected to two physically grounded postulates---the phase-locking (real-amplitude) condition on the chiral condensate ($\overline \psi_R \psi_L \in \mathbb{R}$) and the pure-electric restriction on the Maxwell field---the background geometry is uniquely constrained to belong to the Kastor-Traschen class. 

\section{The Nonlinear Spinor System}
We consider the four-dimensional spacetime with the metric $g_{\mu\nu}$,
and the Maxwell field $F_{\mu\nu}$, subject to the cosmological Einstein-Maxwell
equations:
\begin{align}
\label{eq:einstein}  R_{\mu\nu}-\dfrac{R}{2}g_{\mu\nu}+\varLambda g_{\mu\nu}
  &=-2F_{\mu\lambda}F_\nu{}^\lambda+\dfrac{1}{2}F_{\lambda\rho}F^{\lambda\rho}g_{\mu\nu}\\
\label{eq:maxwell}  F_{\mu\nu}&=\partial_\mu A_\nu-\partial_\nu A_\mu,
\end{align}
with the positive cosmological constant $\varLambda>0$.
We adopt the $(+,-,-,-)$ metric signature,
and the Clifford algebra relation is given by
\begin{align*}
  \gamma^a  \gamma^b+  \gamma^b  \gamma^a=2\eta^{ab},
\end{align*}
where $\eta_{ab}=\operatorname{diag}(1,-1,-1,-1)$. 
The super-covariant derivative operator $D_\mu$ for the Dirac spinor $\psi$
is defined by
\begin{align*}
  D_\mu\psi=\nabla_\mu\psi-\dfrac{1}{4}F_{\nu\lambda}\gamma^\nu
  \gamma^\lambda \gamma_\mu\psi,
\end{align*}
where $\gamma^\mu$ is defined in terms of tetrad $\{e_a{}^\mu\}_{a=0,1,2,3}$
as $\gamma^\mu =e_a{}^\mu \gamma^a$, and the
spin connection $\nabla_\mu$ is given by
\begin{align*}
  \nabla_\mu\psi=\partial_\mu\psi+\dfrac{1}{4}\varGamma_{ab\mu}
  \gamma^a\gamma^b\psi,
\end{align*}
in terms of the Ricci rotation coefficients $\varGamma_{ab\mu}=e_a{}^\nu\nabla_\mu e_{b\nu}$.

Now, we state the following theorem:
{\flushleft{\bf Theorem}.}
{\it
Let a spacetime  admit a Dirac spinor $\psi$
satisfying Eq.~(\ref{eq:scc}) for a positive constant $H$.
If the hermitian product $W=\overline\psi_R\psi_L$
of the left- and the right-handed components is a real-valued
 function without zero,
and the pure-electric condition
\begin{align*}
  \overline\psi\gamma_{[a}\psi F_{bc]}=0
\end{align*}
is satisfied, then the spacetime metric and the Maxwell field
are restricted to the following
Kastor-Traschen form
\begin{align*}
  ds^2&=U^{-2}dt^2-U^2(dx^2+dy^2+dz^2),\\
  F&= -U^{-2}dU\wedge dt,\\
\end{align*}
where     $U=Ht+V(x,y,z)$.
}

\vspace{12pt}
Therefore, the only solution to the cosmological
Einstein-Maxwell equations (\ref{eq:einstein}), (\ref{eq:maxwell})
for $\varLambda=3H^2$ satisfying the above conditions is
the Kastor-Traschen solution, where $V$ is a harmonic function
on $\mathbb{R}^3$.

{\bf Remark.} Crucially, the nonlinear system~(\ref{eq:scc}) automatically implies that the spinor becomes an eigenvector of the super-covariant Dirac operator,
\begin{align*}
  \gamma^\mu D_\mu \psi=H\psi,
\end{align*}
demonstrating that the Hubble parameter $H$ directly emerges as the physical spectrum of the geometry-coupled fermion.

\vspace{12pt}
{\flushleft \it Proof.}
We adopt the chiral representation for the Dirac matrices as
\begin{align*}
  \gamma^0=\begin{pmatrix}
             0&\mathbbm{1}\\
\mathbbm{1}            &0
\end{pmatrix},~~~
  \gamma^i=\begin{pmatrix}
             0&\sigma^i\\
-\sigma^i             &0
\end{pmatrix},~~~(i=1,2,3)
\end{align*}
where $\sigma^i$'s are the Pauli matrices.
We write the Dirac spinor as
\begin{align*}
  \psi=\begin{pmatrix}
    \alpha^A\\\beta_{A'}
    \end{pmatrix}~~~(A,A'=0,1)
\end{align*}
in terms of two-component Weyl spinors $\alpha^A$, $\beta_{A'}$.
Here, the undotted spinor \(\alpha ^{A}\) is left-handed, whereas the dotted spinor \(\beta _{A^{\prime }}\) is right-handed.
The hermitian product of the chiral components of $\psi$ is written as
\begin{align*}
  W=\overline\psi_R\psi_L=\alpha^A\overline\beta_A,
\end{align*}
which is assumed to be a real-valued function.
From Eq. (\ref{eq:scc}),
$\alpha^A$ and $\beta_{A'}$ are subject to 
\begin{align}
\label{eq:scc1}  \nabla_{AA'}\alpha^B+\sqrt{2}\left(\phi_A{}^B-\dfrac{H}{2}\epsilon_A{}^B\right)\beta_{A'}+\dfrac{H}{\sqrt{2}W}(\alpha_A\overline\alpha_{A'}
  +\overline\beta_A\beta_{A'})\alpha^B&=0,\\
\label{eq:scc2}  \nabla_{AA'}\beta_{B'}-\sqrt{2}\left(\overline\phi_{A'B'}-\dfrac{H}{2}\epsilon_{A'B'}\right)\alpha_A+\dfrac{H}{\sqrt{2}W}(\alpha_A\overline\alpha_{A'}
  +\overline\beta_A\beta_{A'})\beta_{B'}&=0,
\end{align}
where $\epsilon_{AB}$ 
denotes the antisymmetric spinor
($\epsilon_{01}=\epsilon^{01}=1$),
and
$\phi_{AB}$ denotes the symmetric spinor arising from
the irreducible decomposition of the Maxwell field.

Consider the mixed component spinors 
\begin{align*}
  L_{AA'}&=\alpha_A\overline\alpha_{A'},~~~
  N_{AA'}=\overline\beta_A\beta_{A'},\\
  M_{AA'}&=\alpha_A\beta_{A'},~~~
\overline  M_{AA'}=\overline\beta_A\overline\alpha_{A'},
\end{align*}
which corresponds to complexified tangent vectors
$L^a$, $N^a$, $M^a$, $\overline M^a$
as
\begin{align*}
  L_{AA'}=\dfrac{1}{\sqrt{2}}\begin{pmatrix}
    L_0-L_3&-L_1-iL_2\\
    -L_1+iL_2&L_0+L_3
    \end{pmatrix}_{AA'}
\end{align*}
etc. Since
\begin{align*}
L^aL_a=  N^aN_a=M^aM_a=0,~~~L^aN_a=-M^a\overline M_a=W^2
\end{align*}
hold, $(L^a,N^a,M^a,\overline M^a)$ is regarded as a prenormalized
Newman-Penrose basis. Hence the metric can be written as
\begin{align*}
  g_{\mu\nu}=2W^{-2}(L_{(\mu}N_{\nu)}-M_{(\mu}\overline M_{\nu)}).
\end{align*}

Noting that the spinor representation of the
current vector  $K^a=\overline\psi\gamma^a\psi$
becomes
\begin{align*}
  K_{AA'}=\sqrt{2}(\alpha_A \overline\alpha_{A'}+\overline\beta_A\beta_{A'})
  =\sqrt{2}(L_{AA'}+N_{AA'}),
\end{align*}
Eqs. (\ref{eq:scc1}), (\ref{eq:scc2}) imply
\begin{align}
\label{eq:dk}  \nabla_\mu K_\nu-  \nabla_\nu K_\mu
  &=-2W F_{\mu\nu},\\
\label{eq:gckv}  \nabla_\mu K_\nu+  \nabla_\nu K_\mu
&  =4HW g_{\mu\nu}-2HW^{-1} K_\mu K_\nu.
\end{align}
The pure-electric condition $K\wedge F=0$, combined with Eq.~(\ref{eq:dk})
implies that the vector field $K^\mu$ is hypersurface orthogonal.
Hence a local coordinate system $(x^0,x^1,x^2,x^3)$ can be chosen
such that $K^\mu=(1,0,0,0)$ and that $K^\mu$ is orthogonal
to hypersuface $x^0=0$. Then, $g_{0\mu}=K_\mu$ and
$g_{00}=4W^2$ hold.  In this coordinate system,
The Eq.~(\ref{eq:gckv}) becomes
\begin{align*}
  \partial_0g_{00}&=-16HW^3,\\
  \partial_0g_{0i}&=-2HW^{-1}(g_{0i})^2+4HWg_{0i},\\
  \partial_0g_{ij}&=-2HW^{-1}g_{0i}g_{0j}+4HWg_{ij}.
\end{align*}
Solving these under the initial conditions $g_{00}=4/V(x^k)^2$,
$g_{0i}=0$, $g_{ij}=h_{ij}(x^k)$ at $x^0=0$,
\begin{align*}
  W^{-1}&=2Hx^0+V(x^k),~~~
  g_{0i}=0,~~~g_{ij}=W^2h_{ij}(x^k)
  \end{align*}
  are obtained.

  On the other hand, Eq.~(\ref{eq:scc1}), (\ref{eq:scc2}) also imply
  \begin{align*}
    \nabla_{[\mu}(L-N)_{\nu]}= \nabla_{[\mu}M_{\nu]}=0,
  \end{align*}
so that $L-N$ and $M$ are closed forms subject to
$K^a(L_a-N_a)=K^aM_a=0$.
Hence local functions $x$, $y$, $z$ can be chosen which depend only on
$x^k$, such that
\begin{align*}
  L-N=\sqrt{2}dz,~~~M=\dfrac{1}{\sqrt{2}}(dx-idy)
\end{align*}
hold. Under the new coordinates $t=2x^0$, $x$, $y$, $z$,
the metric can be written as
\begin{align*}
  ds^2=U^{-2}dt^2-U^2(dx^2+dy^2+dz^2),
  \end{align*}
  where $U=Ht+V(x,y,z)$.
  Furthremore, Eq.~(\ref{eq:dk}) leads to
  \begin{align*}
    F=-U^{-2}dU\wedge dt.
      \end{align*}
\hfill \qedsymbol

\section{Discussion}
We have demonstrated that a dynamical multi-black-hole spacetime
naturally emerges from a purely intrinsic, nonlinear spinor system.
By deforming the standard super-covariant derivative
with a spinor current interaction,
the geometry of an expanding universe becomes rigidly locked.
This result represents a conceptual leap from the classic static classifications of BPS vacua:
the cosmic expansion is not merely a background configuration,
but is actively enforced by the matter sector through
a geometric eigenvalue problem.
Our work thus establishes a cosmic uniqueness theorem for
the Kastor-Traschen geometry,
revealing that the exact solution to
the cosmological Einstein-Maxwell equations is uniquely hardwired into a deformed nonlinear Dirac system.

Our formulation establishes a novel dictionary between cosmology
and quantum geometry.
Crucially, the Hubble constant $H$, representing
the macroscopic rate of cosmic expansion,
emerges fundamentally as the spectrum of the underlying
super-covariant Dirac operator.
The conceptual breakthrough of our proof lies in demonstrating that
the pure-electric condition is equivalent to
the hypersurface orthogonality of the resulting spinor current.
This equivalence bridges the gap between microscopic fermion
self-interactions and the time function in dynamic black hole spacetimes.

It is worth emphasizing that the specific structure of
the nonlinear fermion self-interaction in Eq.~\eqref{eq:scc} was
deduced via a reverse-engineering of the Kastor-Traschen geometry.
The lack of Killing spinors in dynamical spacetimes poses a well-known
challenge for their characterization.
Our results demonstrate that the nonlinear term proportional
to $H/W^2$ provides precisely the unique geometric correction required
for this setting.
Given this construction, our findings strongly suggest the existence of a deeper,
yet undiscovered symmetry in cosmological supergravity.
This framework opens up a new avenue for exploring
the hidden integrability of non-stationary solutions in general relativity,
and we hope our work stimulates
further exploration into the spectral origin of spacetime dynamics.

\end{document}